\documentclass{article}

\usepackage{amsmath,amssymb,amsfonts}
\usepackage{bm}
\usepackage{color}
\usepackage{fancybox}
\usepackage{graphicx}
\usepackage{a4wide}

\title{Boundary effects on the lattice/continuum correspondence: the spin-$1/2$ XXZ chain and the sine-Gordon model}
\author{Chihiro Matsui \\[3ex]
{\it Department of Mathematical Informatics, The University of Tokyo} \\
{\it 7-3-1 Hongo, Bunkyo-ku, Tokyo 113-8656, Japan}}

\begin{document}
\maketitle

\abstract{
We derived the corresponding boundary condition on Fermi fields to the spin-$1/2$ Heisenberg 
chain with boundary magnetic fields. In order to obtain the correct boundary condition from 
the variation of the action at the edges, we carefully treat the oscillating terms which emerge 
as a result of the chiral decomposition of fermions and do not contribute to the bulk Lagrangian. 
The obtained result is checked by compared with the exact result derived from the Bethe ansatz, 
by considering the mode expansion of fermions on the light-cone coordinates. We also give 
the spin-wave interpretation to the emergence of boundary bound states. 
}

\section{Introduction} \label{sec:int}
There have been many discussions about the field theoretic treatment of 
spin chains \cite{bib:H82, bib:H83, bib:A90, bib:T90}. On of the most widely used method is 
the bosonization. 
The strategy goes on first to express spin operators in terms of 
fermions and then decomposing them into left-movers and right-movers at the Fermi surface. 
Usually, the linear dispersion relation is assumed since we are interested in the low-energy regime. 
This linearlization of the dispersion relation makes it possible to bosonize the fermionic theory. 
The dispersion relations of both particles and holes are 
linear-like at the Fermi surface. As a result, the density operator acts as a boson operator. 
Thus, the gapless spin chain admits the boson description at the low-energy regime. 

In spite of successful progress in analysis of critical exponent and the corresponding conformal 
field theory \cite{bib:A90, bib:A86, bib:EA92}, at the 
same time with the available expressions of the {\it continuum} $SU(2)$ spin operators 
in terms of Fermi fields \cite{bib:LP75}, many of discussions have been made only for the periodic boundary
case. Therefore, there still remain the question of {\it what is the corresponding boundary conditions
on the Fermi fields to the boundary magnetic fields}. One also obtains that resolving this problem 
finally proves the statement that the corresponding continuum field theory to the Heisenberg spin chain
{\it with boundary magnetic fields} is the sine-Gordon model {\it with Dirichlet boundary conditions}. 
Also, taking the continuum limit of the spin chain sheds light on why a spin chain consisting of 
an even (odd) number of sites admits only an even (odd) number of excitation particles in its
effective field theory. Although there exist several papers \cite{bib:AKL95, bib:HSWY97, bib:CSS03} which discussed how the 
scattering process of the fermionic theory can be read off from the bosonic theory, the correspondence to the spin chain 
cannot be fully understood from them, since they lack the important effect of discreteness coming 
from whether the lattice consists of an even number of sites or an odd number of sites, including the above restriction on excitations. 

The aim of this paper is to derive the valid boundary condition on Fermi fields through
the bosonization of the Heisenberg spin chain with boundary magnetic fields. In the 
next section, we briefly review the bosonization method. In the third section, the valid boundary 
conditions are derived through the variation method. The available 
definition of the continuum $SU(2)$ spin operators are also reconsidered. 
Then the forth section is devoted to checking 
the compatibility of the derived boundary conditions with the known exact result from 
the scattering theory. In the last section, we give a concluding remarks and future problems.

\section{Bosonization} \label{sec:boson}
The idea of the bosonization originates from boson-like particle-hole excitations, which is 
specifically obtained in one-dimensional electron systems. Also, the Heisenberg spin chain 
is a famous example which admits a bosonized form, due to its Luttinger-liquid behavior. 
The Heisenberg spin chain is defined by the following Hamiltonian: 
\begin{equation} \label{hamiltonian-xxz}
 H = \frac{J}{2}\left[
		  \sum_{n=1}^{\ell-1} 
		  \left(S_n^+ S_{n+1}^- + S_n^- S_{n+1}^+ + 4\Delta S_n^z S_{n+1}^z\right)
		  + 2h_+ S_1^z + 2h_- S_\ell^z
       \right]. 
\end{equation}
On both ends of the spin chain, the magnetic fields are imposed. 
The parameter $\Delta = J_z/2J = \cos\gamma$ determines the anisotropy of the model. For $J^z = 0$ 
(the XY model), the effective theory results in the free fermion. We focus on the gapless regime $0 \leq J^z \leq 1$ throughout
this paper. 

In order to apply the bosonization method to the spin chain, one first needs to express the spin operators 
in terms of Fermi fields. This is achieved by the well-known Jordan-Wigner transformation, 
by regarding $S^-$ with an annihilation operator and $S^+$ with a creation operator, but with 
the string operator in order to correctly reproduce the $SU(2)$ algebra: 
\begin{align}
 &S_n^- = \psi_n \exp\left[i\pi \sum_{j=1}^{n-1} \psi_j^{\dag}\psi_j\right], 
 \quad
 S_n^+ = \exp\left[-i\pi \sum_{j=1}^{n-1} \psi_j^{\dag}\psi_j\right] \psi_n^{\dag}, 
 \\
 &S_n^z = \psi_n^{\dag} \psi_n. 
\end{align}
Here the subscript $n$ denotes the site number which each operator acts on. 
Then we decompose a Fermi field into a left-mover and a right-mover in the vicinity of the Fermi surface: 
\begin{equation} \label{fermi-d}
 \psi_n \sim i^n \psi_{{\rm L},n} + (-i)^n \psi_{{\rm R},n}. 
\end{equation} 
From the relation above, one may notice that the characteristic feature of the discreteness emerges 
through the site-dependent oscillating coefficients. The effect 
of discreteness is indeed very important for boundary conditions. For instance, if the 
periodic spin chain consists of an even number of sites, the corresponding field theory has 
periodic boundaries as well, while for the spin chain consisting of an odd number of sites, 
 the anti-periodic boundary condition is imposed on the effective field theory. 

Taking the continuum limit of (\ref{fermi-d}) results in 
\begin{equation} \label{lat-c}
 \frac{\psi(x)}{\sqrt{a}} \simeq e^{i(\pi/2a)x} \psi_{\rm L}(x) + e^{-i(\pi/2a)x} \psi_{\rm R}(x), 
\end{equation}
where $a$ is a lattice constant. 
Assuming the linear dispersion near the Fermi surface, the XY part of (\ref{hamiltonian-xxz}) 
is described by the free fermion theory: 
\begin{equation} \label{hamiltonian-free}
 H_{\rm free} = -iv \int_0^L dx\; 
  \left[\psi^{\dag}_{\rm L} \frac{d}{dx} \psi_{\rm L} 
   - \psi^{\dag}_{\rm R} \frac{d}{dx} \psi_{\rm R}\right], 
\end{equation}
where $v$ is the velocity of light given by $v = Ja$. 
One has to be careful that the cross terms such as $\psi^{\dag}_{\rm L} \psi_{\rm R}$ do not 
show up since they cancel out due to the oscillating coefficients (\ref{fermi-d}) by summing up 
over all sites. 
The important key point in dealing with open boundary cases is to keep these oscillating terms, 
as we discuss in Section \ref{sec:bound}. 

In order to deal with the $S^z$-interaction term, we first introduce the chiral currents 
$J_{\rm L,R} = :\psi^{\dag}_{\rm L,R} \psi_{\rm L,R}:$. Here we used the normal order $:*:$. 
Then the free part (\ref{hamiltonian-free}) is expressed by these currents as 
\begin{equation}
 H_{\rm free} = \pi Ja \int_0^L dx\; 
  \left[J_{\rm L}(x) J_{\rm L}(x) + J_{\rm R}(x) J_{\rm R}(x)\right]. 
\end{equation}
The $S^z$-interaction term is also written by chiral currents:  
\begin{equation} \label{hamiltonian-int}
 H_{\rm int} = -J_z a \int_0^L dx\; 
  \left[J_{\rm L}^2 + J_{\rm R}^2 + 2(J_{\rm L} J_{\rm R} + J_{\rm R} J_{\rm L})
  - 2 \{(\psi^{\dag}_{\rm L} \psi_{\rm R})^2 + (\psi^{\dag}_{\rm R} \psi_{\rm L})^2\}\right]. 
\end{equation}

By using the know bosonization formula \cite{bib:A90}: 
\begin{equation} \label{bosonization}
\begin{split}
 &J_{\rm L} = i\frac{1}{\sqrt{4\pi}} \partial_z \phi_{\rm L}, 
  \quad 
  J_{\rm R} = -i\frac{1}{\sqrt{4\pi}} \partial_{\bar{z}} \phi_{\rm R}, 
 \\
 &\psi_{\rm L} = e^{-i \sqrt{4\pi} \phi_{\rm L}},
 \quad
 \psi_{\rm R} = e^{i\sqrt{4\pi} \phi_{\rm R}}, 
\end{split}
\end{equation}
the first two terms result in the free boson theory and the third term gives rise to
the irrelevant perturbation: 
\begin{equation}
 H_{\rm free} + H_{\rm int}
  =
  \left(J + \frac{J_z}{\pi}\right)a 
  \left[
   2 \left(1 - \frac{2J_z}{\pi J}\right) \int_0^L dx\; \partial_{\mu} \phi\, \partial^{\mu} \phi
   + \frac{4J_z}{\pi J} \int_0^L dx\; \cos(2 \sqrt{4\pi} \phi)
  \right], 
\end{equation}
where $\phi = \phi_{\rm L} + \phi_{\rm R}$ and 
$\partial_{\mu}\phi\, \partial^{\mu}\phi = -\partial_t\phi\, \partial^t\phi + \partial_x\phi\, \partial^x\phi$. 
Here $z$ is a holomorphic coordinate $z = i(t-x)/2$ while $\bar{z}$ is an anti-holomorphic 
coordinate $\bar{z} = i(t+x)/2$. 
Hence, the velocity of light is renormalized as 
\begin{equation}
 v = Ja\; \to\; v = (J + J_z/\pi)a. 
\end{equation}
Using the following replacement: 
\begin{equation}
 \sqrt{4\pi} \phi\; \to\; \phi/R, 
\end{equation}
where the compactification radius $R = [(1 - 2J_z/\pi J)/4\pi]^{1/2}$, we obtain the sine-Gordon Lagrangian: 
\begin{equation}
 L = \frac{1}{2} \int_0^L dx\;
  \left[
   \partial_{\mu}\phi\, \partial^{\mu}\phi - \frac{m^2}{\widehat{\beta}^2} \cos(\widehat{\beta}\phi)
  \right]
\end{equation}
with a bare coupling constant $\widehat{\beta} = R^{-1}$ and $m$ is a soliton mass 
up to the first order of $J_z/J$.

\section{Boundary conditions} \label{sec:bound}
Now our question is how the boundary conditions on the Fermi fields come out. The straightforward way
to derive the boundary conditions is to take the variation of the theory on both ends. 
However, just by naively taking the variation of the fremionic representation of the spin chain 
does not give correct boundary 
conditions. We have got rid of the oscillating terms in the local action due to the summation
all over the sites, which play the important role for determining boundary conditions.

\subsection{XY case with boundary magnetic fields}
For simplicity, we start from the boundary effects on the XY model, whose low-energy effective 
field theory is the free fermion. 
The bulk part of the XY-chain Hamiltonian looks in terms of fermions like  
\begin{equation} \label{hamiltonian-free-f}
\begin{split}
 H_{\rm free} &= \frac{J}{2} \sum_{n=1}^{\ell-1} 
  \left[\psi^{\dag}_{n+1} \psi_n + \psi^{\dag}_n \psi_{n+1}\right] 
 \\
 &= \frac{iJ}{2} \sum_{{\rm odd}\, n} 
 \left[(\psi^{\dag}_{{\rm L},n} - \psi^{\dag}_{{\rm R},n}) (\psi_{{\rm L},n+1} + \psi_{{\rm R},n+1})
 - (\psi^{\dag}_{{\rm L},n+1} + \psi^{\dag}_{{\rm R},n+1}) (\psi_{{\rm L},n} - \psi_{{\rm R},n})\right]
 \\
 &+ \frac{iJ}{2} \sum_{{\rm even}\, n} 
 \left[(\psi^{\dag}_{{\rm L},n} + \psi^{\dag}_{{\rm R},n}) (\psi_{{\rm L},n+1} - \psi_{{\rm R},n+1})
 - (\psi^{\dag}_{{\rm L},n+1} - \psi^{\dag}_{{\rm R},n+1}) (\psi_{{\rm L},n} + \psi_{{\rm R},n})\right]. 
\end{split}
\end{equation}
Although the cross terms such as $\psi^{\dag}_{\rm L} \psi_{\rm R}$, which have oscillating 
coefficients, cancel out after the summation and do not contribute to the bulk Hamiltonian, 
we need to carefully deal with those terms 
in considering the boundary conditions. By using the difference form for small $a$, for instance such as 
\begin{equation} \label{difference}
 \psi_{{\rm L},n+1} = \psi_{{\rm L},n} + a \frac{\psi_{{\rm L},n+1} - \psi_{{\rm L},n}}{a}, 
\end{equation}
we obtain the leading term of the diagonal part disappears but the subleading 
term remains finite, which results in the free-fermion Hamiltonian. 
Let us remark that the last term of (\ref{difference}) is the order $a$, 
since it becomes $a \partial_x \psi_{\rm L}(x,t)|_{x=an}$ in the continuum limit $a \to 0$. 
On the other hand, 
the leading term of {\it the cross part remains finite at the one-site level}. Of course, 
these terms do not emerge after the summation {\it i.e.} the integration in the continuum limit. 
However, this cross part is crucially important for the boundary conditions. 

The boundary terms in (\ref{hamiltonian-xxz}) is expressed by chiral fermions as 
\begin{equation} \label{hamiltonian-b}
 H_{\rm b} = h_+J (\psi^{\dag}_{{\rm L},0} + \psi^{\dag}_{{\rm R},0})(\psi_{{\rm L},0} + \psi_{{\rm R},0}) 
  + h_-J (i^{\ell}\psi_{{\rm L},\ell} + (-i)^{\ell}\psi_{{\rm R},\ell})^{\dag}(i^{\ell}\psi_{{\rm L},\ell} + (-i)^{\ell}\psi_{{\rm R},\ell}). 
\end{equation}
After converting the Hamiltonian (\ref{hamiltonian-free-f}) under the boundary conditions 
(\ref{hamiltonian-b}) to the Lagrangian, 
we apply the discrete-analogue of the variation method \cite{bib:FM98, bib:F99, bib:F01} to the edges of the Lagrangian 
and then obtain the stationary condition with respect to $\psi^{\dag}_{\rm L}$ and 
$\psi^{\dag}_{\rm R}$ for $n=0$ as 
\begin{equation} \label{0site_cond}
 i(\psi_{{\rm L},0} - \psi_{{\rm R},0}) - h_+ (\psi_{{\rm L},0} + \psi_{{\rm R},0}) =0, 
\end{equation} 
whereas that for $n=\ell$ as 
\begin{equation} \label{Lsite_cond}
\begin{split}
 &i(\psi_{{\rm L},\ell} - \psi_{{\rm R},\ell}) - h_- (\psi_{{\rm L},\ell} + \psi_{{\rm R},\ell}) = 0
 \qquad \text{for even }\ell, \\
 &i(\psi_{{\rm L},\ell} + \psi_{{\rm R},\ell}) - h_- (\psi_{{\rm L},\ell} - \psi_{{\rm R},\ell}) = 0
 \qquad \text{for odd }\ell. 
\end{split}
\end{equation}
By taking $h_{\pm} \to 0$ in (\ref{0site_cond}) and (\ref{Lsite_cond}), we obtain 
the boundary conditions for the zero-boundary-field case: 
\begin{equation} \label{zero-b-cond}
\begin{split}
 &\psi_{{\rm L},0} + \psi_{{\rm R},0} = 0, 
  \quad
 \psi_{{\rm L},\ell} + \psi_{{\rm R},\ell} = 0
 \qquad \text{for even }\ell, \\
 &\psi_{{\rm L},0} + \psi_{{\rm R},0} = 0, 
  \quad
 \psi_{{\rm L},\ell} - \psi_{{\rm R},\ell} = 0 
 \qquad \text{for odd }\ell. 
\end{split}
\end{equation}
This boundary condition was treated first in \cite{bib:EA92} by introducing phantom sites at $n=0,\ell+1$, where fermions vanish. 

Here let us remark the continuum representations of the spin operators by Fermi fields. 
Although we proceed our discussion on the discrete system first and then take the continuum limit 
at the end when we compare our results with those obtained from the $S$-matrix theory, 
the continuum version of the Hamiltonians (\ref{hamiltonian-free-f}) and (\ref{hamiltonian-b}) 
can directly be obtained by replacing the spin operators with continuum analogues: 
\begin{equation} \label{spin-cont}
\begin{split}
 &\sqrt{2a^{-1}}\,S^-(x) = [e^{i (\pi/2a) x} \psi_{\rm L}(x) + e^{-i (\pi/2a) x} \psi_{\rm R}(x)] e^{-N(x)}, 
  \quad
  S^+(x) = [S^-(x)]^{\dag}, 
 \\
 &a^{-1}\,S^z(x) = \rho_{\rm L}(x) + \rho_{\rm R}(x)
  + e^{i (\pi/a) x} \psi^{\dag}_{\rm L} \psi_{\rm R}(x) + e^{-i (\pi/a) x} \psi^{\dag}_{\rm R} \psi_{\rm L}(x), 
\end{split}
\end{equation}
where 
\begin{equation}
 N(x) = i\pi \int_0^{x-a/2} dy\;
  \left[\rho_{\rm L}(y) + \rho_{\rm R}(y)\right]. 
\end{equation}
The density operator $\rho_{\rm L,R}$ is defined by 
$\rho_{\rm L/R}(x) = \psi^{\dag}_{\rm L/R}(x) \psi_{\rm L/R}(x) - 1/2 = :\psi^{\dag}_{\rm L/R}(x) \psi_{\rm L/R}(x):$, 
where $:*:$ is the normal order. 
The above continuum representations of the spin operators are quite similar to what have been 
introduced in \cite{bib:LP75} but for the site-dependent phase shifts, which play an important 
role for the boundary conditions. 
The algebraic relations for $S^{\pm,z}$: 
\begin{equation}
 [S^+(x),\, S^-(x')] = 2S^z(x) \delta(x-x'), 
  \quad
  [S^{\pm}(x),\, S^z(x')] = \mp S^{\pm}(x) \delta(x-x'), 
\end{equation}
which ensure the $SU(2)$-symmetry of the XXZ chain, can be checked by direct calculation.

Using the bosonization formula (\ref{bosonization}) 
with $R=1/\sqrt{4\pi}$ for the free theory, the boundary conditions (\ref{0site_cond}) 
and (\ref{Lsite_cond}) read 
\begin{equation}
 \phi(0) = \pi R, 
  \quad 
  \phi(\ell) = 2\pi R \left(\widehat{n}+\frac{1}{2}\right), 
\end{equation}
where $\widehat{n}$ is an integer for even $\ell$ and a half-integer for odd $\widehat{n}$. Thus, our procedure 
correctly reproduces the result obtained in \cite{bib:EA92}.

\subsection{XXZ case with boundary magnetic fields}
Now we proceed to the interacting system. The $S^z$-interaction term (\ref{hamiltonian-int}) is 
written by using the normal order as 
\begin{equation}
\begin{split}
 H_{\rm int} = \sum_{n=1}^{\ell-1} 
  &:\psi^{\dag}_n \psi_n:\,:\psi^{\dag}_{n+1} \psi_{n+1}:
 \\
 = \sum_{n=1}^{\ell-1}
 &\Big[
 \Big(\rho_{{\rm L},n} + \rho_{{\rm R},n} + (-1)^n \psi^{\dag}_{{\rm L},n} \psi_{{\rm R},n} + (-1)^n \psi^{\dag}_{{\rm R},n} \psi_{{\rm L},n}\Big) 
 \\
 \times 
 &\Big(\rho_{{\rm L},n+1} + \rho_{{\rm R},n+1} + (-1)^{n+1} \psi^{\dag}_{{\rm L},n+1} \psi_{{\rm R},n+1} + (-1)^{n+1} \psi^{\dag}_{{\rm R},n+1} \psi_{{\rm L},n+1}\Big) 
 \Big]. 
\end{split} 
\end{equation}
Thus we obtain the leading term disappears at the one-site level, which implies 
the existence of the interaction does not affect the boundary conditions.

Let us remark here that the discussion above is valid only for small $J_z/J$, 
as referred many times \cite{bib:A90, bib:LP75}. The reason is partly because we linearly expand 
the Fermi fields at the vicinity of the free Fermi surface, which is no more the genuine 
Fermi surface under the presence of interactions. We will see the example of this problem
in checking the accuracy of the boundary conditions (\ref{0site_cond}) and (\ref{Lsite_cond}).

\section{Check from the $S$-matrix theory} \label{sec:s-mat}
As the XXZ model is an integrable system, the exact $S$-matrix and the reflection matrix
have been derived \cite{bib:ZZ78, bib:GZ94, bib:FS94}. The reflection of a soliton 
$A^{\dag}_+(\theta)$
(and an anti-soliton $A^{\dag}_-(\theta)$) is written in forms of 
\begin{equation} \label{ref-rel}
\begin{split}
 &A^{\dag}_+(\theta) B = P^+(\theta) A^{\dag}_+(-\theta) B + Q^+(\theta) A^{\dag}_-(-\theta) B, 
  \\
 &A^{\dag}_-(\theta) B = Q^-(\theta) A^{\dag}_+(-\theta) B + P^-(\theta) A^{\dag}_-(-\theta) B, 
\end{split}
\end{equation}
where $B$ is the boundary operator. 
Corresponding to the boundary magnetic fields, the diagonal solution of the reflection relation 
\cite{bib:S88} has been obtained as 
\begin{equation} \label{ref-th}
\begin{split}
 &P^+(\theta) = \left(\theta + \frac{i\pi H_+}{2}\right), 
  \quad
  P^-(\theta) = \left(\frac{i\pi H_-}{2} - \theta\right), \\
 &Q^+(\theta) = Q^-(\theta) = 0, 
\end{split}
\end{equation}
where boundary magnetic fields are related to $H_{\pm}$ by 
$h_{\pm} = \sin\gamma \cot(\gamma H_{\pm}/2)$.

In order to obtain the reflection amplitudes from the fermionic description, we consider 
the mode expansion of the fermions on the light-cone basis: 
\begin{equation}
\begin{split}
 &\psi_{\rm L} = \sqrt{m} \int_{-\infty}^{\infty} \frac{d\theta}{2\pi i} e^{\theta/2}
 \left[A_-(\theta) e^{-m(ze^{\theta} + \bar{z}e^{-\theta})}
 - A^{\dag}_+(\theta) e^{m(ze^{\theta} + \bar{z}e^{-\theta})}\right], 
 \\
 &\psi_{\rm R} = -i\sqrt{m} \int_{-\infty}^{\infty} \frac{d\theta}{2\pi i} e^{-\theta/2}
 \left[A_-(\theta) e^{-m(ze^{\theta} + \bar{z}e^{-\theta})}
 + A^{\dag}_+(\theta) e^{m(ze^{\theta} + \bar{z}e^{-\theta})}\right], 
\end{split}
\end{equation}
where $m$ is a soliton mass we obtained in Section \ref{sec:boson}. 
By substituting these mode expansions and their conjugate  into the derived boundary 
conditions (\ref{0site_cond}) and (\ref{Lsite_cond}), we obtain the reflection relations 
are written in the forms of (\ref{ref-rel}) if we choose $P^{\pm}(\theta)$ as 
\begin{equation} \label{ref-mat}
 P^+(\theta) = i \frac{(1-h_+^2)\cosh\theta - 2ih_+\sinh\theta}{(1+h_+^2)\sinh\theta - 2ih_+}, 
  \quad
  P^-(\theta) = i \frac{(1-h_-^2)\cosh\theta + 2ih_-\sinh\theta}{(1+h_-^2)\sinh\theta - 2ih_-}. 
\end{equation}
These amplitudes indeed coincide with (\ref{ref-th}) up to an overall factor, if the relation 
$h_{\pm} = \tan(\pi H_{\pm}/4)$ holds, which is obtained as the Taylor expansion around 
$\gamma \sim \pi/2$ of $h_{\pm} = \sin\gamma \cot(\gamma H_{\pm}/2)$.

Now we briefly refer to the corresponding boundary condition on the sine-Gordon model. 
As we already obtained, the sine-Gordon model is realized through the bosonization of 
the XXZ model. The action of the sine-Gordon model with integrable boundaries is written 
as 
\begin{equation} \label{action-sg}
\begin{split}
 S = &\frac{1}{2}\int_0^L dx\,\int dt\;
  \left(\partial_{\mu} \phi\, \partial^{\mu} \phi + \frac{m^2}{\widehat{\beta}^2} \cos(\widehat{\beta} \phi)\right)
 \\
  &- g \int dt\; \cos\left(\frac{\widehat{\beta} (\phi - \phi_+)}{2}\right)
  - g \int dt\; \cos\left(\frac{\widehat{\beta} (\phi - \phi_-)}{2}\right). 
\end{split}
\end{equation}
The stationary conditions obtained from the variation of the action at the both edges 
have been obtained \cite{bib:AKL95} as 
\begin{equation} \label{b-th}
\begin{split}
 &\psi_{\rm L} + e^{i(\mp\phi_{\pm}/R - \sigma)} \psi^{\dag}_{\rm R} - e^{-i\sigma} \psi_{\rm R} = 0, 
 \\
 &i\partial_t (\psi^{\dag}_{\rm L} - e^{i\sigma} \psi_{\rm R}) - g^2 (\psi^{\dag}_{\rm R} e^{\pm i\phi_{\pm}/R} - \psi^{\dag}_{\rm L}) = 0, 
 \\
 &i\partial_t (\psi_{\rm L} - e^{-i\sigma} \psi^{\dag}_{\rm R}) + g^2 (\psi_{\rm L} - e^{\mp i\phi_{\pm}/R} \psi_{\rm R}) = 0, 
\end{split}
\end{equation}
where $\sigma$ is a free parameter. 
By comparing these conditions with (\ref{0site_cond}) and (\ref{Lsite_cond}), we obtain 
that our conditions are realized by taking $g \to \infty$ limit in (\ref{b-th}). One may 
notice that the $g \to \infty$ limit gives rise to {\it the Dirichlet boundary conditions on 
the sine-Gordon model} (\ref{action-sg}). 
The relation between the boundary parameters of the spin chain and the sine-Gordon model 
are then given by 
\begin{equation} \label{spin-boson}
\begin{split}
 &R^{-1}\phi_+ = i \ln \frac{i - h_+}{i + h_+},
  \quad
 R^{-1}\phi_- = i \ln \frac{i + h_-}{i - h_-}
 \qquad \text{for even }\ell, 
 \\
 &R^{-1}\phi_+ = i \ln \frac{i - h_+}{i + h_+},
  \quad
 R^{-1}\phi_- = i \ln \frac{i + h_-}{i - h_-} + \pi
 \qquad \text{for odd }\ell. 
\end{split}
\end{equation}
However, these relations do not coincides with what have been derived through 
the nonlinear integral equations \cite{bib:GZ94, bib:FS94}. As the bosonization 
method provides the correct result only up to the first order of $J_z/J$, 
our results (\ref{spin-boson}) coincides with the exact result: 
\begin{equation}
 R^{-1}\phi_{\pm} = \pm i\ln\frac{i\sin\gamma + h_{\pm}}{i\sin\gamma - h_{\pm}}
\end{equation}
for $\gamma$ close to $\pi/2$. 

One interesting problem on the correspondence between the XXZ chain and the sine-Gordon 
model with the Dirichlet boundaries has been raised in \cite{bib:ABPR08} that the winding number 
depends on the length of the spin chain {\it i.e.} whether the spin chain consists of 
an even number of sites or an odd number of sites. 
This problem is clearly understood from the relations (\ref{spin-boson}). As the conformal 
dimension is given by 
$\Delta = [(\phi_- - \phi_+)/\sqrt{\pi} + \widehat{n}R]^2/2$, 
where the winding number is given by 
\begin{equation}
 \widehat{n}_{\rm mod} = \widehat{n} 
  + \left\lfloor \frac{\phi_+ - \phi_-}{\sqrt{\pi}} \right\rfloor, 
\end{equation}
for $\lfloor * \rfloor$ represents the integer part of $*$, 
the relations (\ref{spin-boson}) implies that $\widehat{n}_{\rm mod}$ takes 
an integer for even $\ell$ and a half-integer for odd $\ell$. 
Thus, a winding number takes either an integer value or a half-integer value depending 
on the system length. 

Moreover, we need to take care of {\it the physical strip} in the expressions (\ref{spin-boson}). 
From the analysis of boundary bound states \cite{bib:GZ94}, rapidities of boundary 
bound states are encoded by the poles in the reflection matrix, and subsequently, 
the boundary bound states emerge if those poles locate in the physical strip $[0,\pi/2)$. 
Our reflection amplitudes (\ref{ref-mat}) have poles in the physical strip and its 
reflection $(-\pi/2,0]$ for $h_{\pm} < 1$, {\it i.e} $R^{-1} \phi_{\pm} \in (\pi/2,\pi]$ 
or $R^{-1} \phi_{\pm} \in [-\pi,-\pi/2)$. 
Thus, the relations (\ref{spin-boson}) are to be modified as 
\begin{equation} \label{spin-boson*}
\begin{split}
 &R^{-1}\phi_+ = i \ln \frac{i + h_+}{i - h_+} + \pi\delta_+,
  \quad
 R^{-1}\phi_- = i \ln \frac{i + h_-}{i - h_-} + \pi\delta_-
 \qquad \text{for even }\ell, 
 \\
 &R^{-1}\phi_+ = i \ln \frac{i + h_+}{i - h_+} + \pi\delta_+,
  \quad
 R^{-1}\phi_- = i \ln \frac{i + h_-}{i - h_-} + \pi(1 + \delta_-)
 \qquad \text{for odd }\ell, 
\end{split}
\end{equation}
where $\delta_{\pm} = 1$ for $R^{-1} \phi_{\pm} \leq -\pi/2$ while 
$\delta_{\pm} = -1$ for $R^{-1} \phi_{\pm} \geq \pi/2$, as long as 
$R^{-1} \phi_{\pm}$ is defined in $[-\pi,\pi)$. 

Phenomenologically, the boundary conditions (\ref{0site_cond}) and (\ref{Lsite_cond}) 
are understood as absorption or emission of the phase of fermions at the boundaries. 
Since the wave length of the Fermi field is taken to be the same as the distance between
two sites, the compactified boson has an integer winding number on the even length chain, 
while a half-integer winding number on the odd length chain. 
The relation (\ref{spin-boson*}) indicates that we obtain boundary bound states 
{\it if there is a phase shift more than $\pi/2$ at the edge.}
That means, if the boundary magnetic 
field is strong enough to absorb (emit) more than $\pi/2$-phase, then this strong field
arrests the outermost spins, and then consequently, on the spin chain effectively 
one-site less, the winding number also shifts as well.

\section{Conclusion} \label{sec:concl}
Throughout this paper, we devoted our efforts to reformulate the boundary conditions 
of the low-energy effective field theory of the Heisenberg spin chain with the 
boundary magnetic fields by the fermionic language. In the formulation, we showed 
the site-dependent oscillating terms play an important role, although they do not 
contribute to the bulk part. In order to keep this oscillating terms, we used 
the discrete-analogue of the variation method to derive the boundary conditions, 
and then introduced the new realization of the continuum spin operators.
This fermionic formulation allows us to obtain 
\begin{enumerate}
 \item the corresponding boundary conditions on fermions to the open spin chain, 
either with or without magnetic boundary fields. 
 \item phenomenological interpretation of the effect of the boundary magnetic field 
on the Fermi field as a phase shift brought by the boundary field absorbing or emitting 
the phase of fermions.  
reflect at the boundary. 
 \item realization of the known reflection matrix by considering the mode 
expansion of fermions on the light-cone coordinates, and consequently, boundary value 
dependence of the matrix form which also concerns with the conformal dimension. 
\end{enumerate}

Besides what have been obtained in this paper listed above, our formulation has 
an advantage to understand the excitation structure of the higher-spin systems. 
Recently, a series of papers \cite{bib:HF12, bib:H13} discussed how the supersymmetry arises on 
the spin chain. The idea is to define the supercharge in such a way that changes 
the length of a spin chain by one site \cite{bib:HF12}. This implies, if we add the boundary 
magnetic field strong enough to arrest the outermost spin and thus to make 
the spin chain relevantly one-site less, it is expected that this boundary magnetic 
field somehow behaves like a supercharge on the spin chain. 

On the other hand, it has been 
obtained that, in the spin-$1$ integrable chain, the boundary values affect not 
only on the winding number but also on the sector of the supersymmetry realized 
as its low-energy effective field theory \cite{bib:M14}. The low-energy effective field theory 
of the spin-$1$ integrable chain is the supersymmetric sine-Gordon model \cite{bib:IO93} 
and the Dirichlet boundaries are said to be realized from the boundary magnetic 
field, as a diagonal solution of the reflection relation which keeps a soliton charge, 
as well as the spin-$1/2$ case. 
However, one may notice, in order to characterize the spin expectation value 
at the edges, we need to introduce two kinds of magnetic fields: 
\begin{equation}
 H_{\rm b} = h^{(1)}_1 S^z_1 + h^{(2)}_1 (S^z_1)^2
  + h^{(1)}_{\ell} S^z_{\ell} + h^{(2)}_{\ell} (S^z_{\ell})^2. 
\end{equation}
This extra degree of freedom, emerging in the higher-spin case, is obtained in 
the scattering matrix as {\it the RSOS degree of freedom}, which subsequently 
brings the supersymmetry in its effective field theory. 
Although there exists a paper which deeply studied the RSOS structure in the 
spin chain \cite{bib:R91}, it is still difficult to see this structure in the excitation 
picture of the spin chain, and thus to understand how the supersymmetry comes out. 

There is an old discussion about the effective field theory of the higher-spin 
chain \cite{bib:A86} through the non-abelian bosonization \cite{bib:W84}. In this context, 
The extra degree of freedom in the higher-spin case is understood as {\it the color} 
of the QCD theory \cite{bib:A86*}, which is naturally introduced by the non-abelian 
field representation of the spin operators. 
The color degree of freedom is somehow more understandable, in the context of 
the field theory, than the RSOS degree of freedom, and we expect these two are 
shown to be related to each other by analyzing the scattering structure. 
In the forthcoming paper, we discuss the above problems, including the effect of 
boundary magnetic field on the effective field theory of the higher-spin chain, 
and show how the QCD structure 
of the spin chain relates to the RSOS structure. 

\section*{Acknowledgements}
 This work is supported by JST/Supporting positive activities for female researchers.

\bibliographystyle{unsrt}
\bibliography{reference}

\end{document}